\documentclass{acmart}
\usepackage{booktabs}
\usepackage{tabularx}
\usepackage{float}
\usepackage{multirow}  
\usepackage{algorithm}
\usepackage{algpseudocode}
\AtBeginDocument{%
  }

\setcopyright{acmlicensed}
\copyrightyear{2018}
\acmYear{2018}
\acmDOI{XXXXXXX.XXXXXXX}
\acmConference[Conference acronym 'XX]{Make sure to enter the correct
  conference title from your rights confirmation email}{June 03--05,
  2018}{Woodstock, NY}
\acmISBN{978-1-4503-XXXX-X/2018/06}

\begin{document}

\title{SmartBugBert: BERT-Enhanced Vulnerability Detection for Smart Contract Bytecode}



\author{Jiuyang Bu}
\affiliation{
  \institution{School of Cyberspace Security, Hainan University}
  \city{Haikou}
  \country{China}
  \postcode{570228}
}

\author{Wenkai Li}
\affiliation{
  \institution{School of Cyberspace Security, Hainan University}
  \city{Haikou}
  \country{China}
  \postcode{570228}
}

\author{Zongwei Li}
\affiliation{
  \institution{School of Cyberspace Security, Hainan University}
  \city{Haikou}
  \country{China}
  \postcode{570228}
}

\author{Zeng Zhang}
\affiliation{
  \institution{School of Cyberspace Security, Hainan University}
  \city{Haikou}
  \country{China}
  \postcode{570228}
}

\author{Xiaoqi Li}
\affiliation{
  \institution{School of Cyberspace Security, Hainan University}
  \city{Haikou}
  \country{China}
  \postcode{570228}
}
\email{csxqli@ieee.org}


\renewcommand{\shortauthors}{Jiuyang Bu et al.}

\begin{abstract}
  Smart contracts deployed on blockchain platforms are vulnerable to various security vulnerabilities. However, only a small number of Ethereum contracts have released their source code, so vulnerability detection at the bytecode level is crucial. This paper introduces SmartBugBert, a novel approach that combines BERT-based deep learning with control flow graph (CFG) analysis to detect vulnerabilities directly from bytecode. Our method first decompiles smart contract bytecode into optimized opcode sequences, extracts semantic features using TF-IDF, constructs control flow graphs to capture execution logic, and isolates vulnerable CFG fragments for targeted analysis. By integrating both semantic and structural information through a fine-tuned BERT model and LightGBM classifier, our approach effectively identifies four critical vulnerability types: transaction-ordering, access control, self-destruct, and timestamp dependency vulnerabilities. Experimental evaluation on 6,157 Ethereum smart contracts demonstrates that SmartBugBert achieves 90.62\% precision, 91.76\% recall, and 91.19\% F1-score, significantly outperforming existing detection methods. Ablation studies confirm that the combination of semantic features with CFG information substantially enhances detection performance. Furthermore, our approach maintains efficient detection speed (0.14 seconds per contract), making it practical for large-scale vulnerability assessment.
\end{abstract}




\maketitle

\section{Introduction}
With the continuous expansion of application scenarios, the number of smart contracts deployed on the blockchain shows explosive growth \cite{Zheng_2023_Blockchain-Based}. Due to the irreversibility of blockchain, it is difficult to repair the vulnerabilities of deployed smart contracts \cite{Chen_2020_Survey}. This makes the security of on-chain smart contracts face serious challenges \cite{Xia_2021_Tradea}. In order to verify the correctness of smart contracts and reduce the losses caused by security issues, a method that can efficiently detect smart contract vulnerabilities is essential \cite{Luu_2016_Making}. 

\begin{figure}[h]
  \centering
  \includegraphics[width=\linewidth]{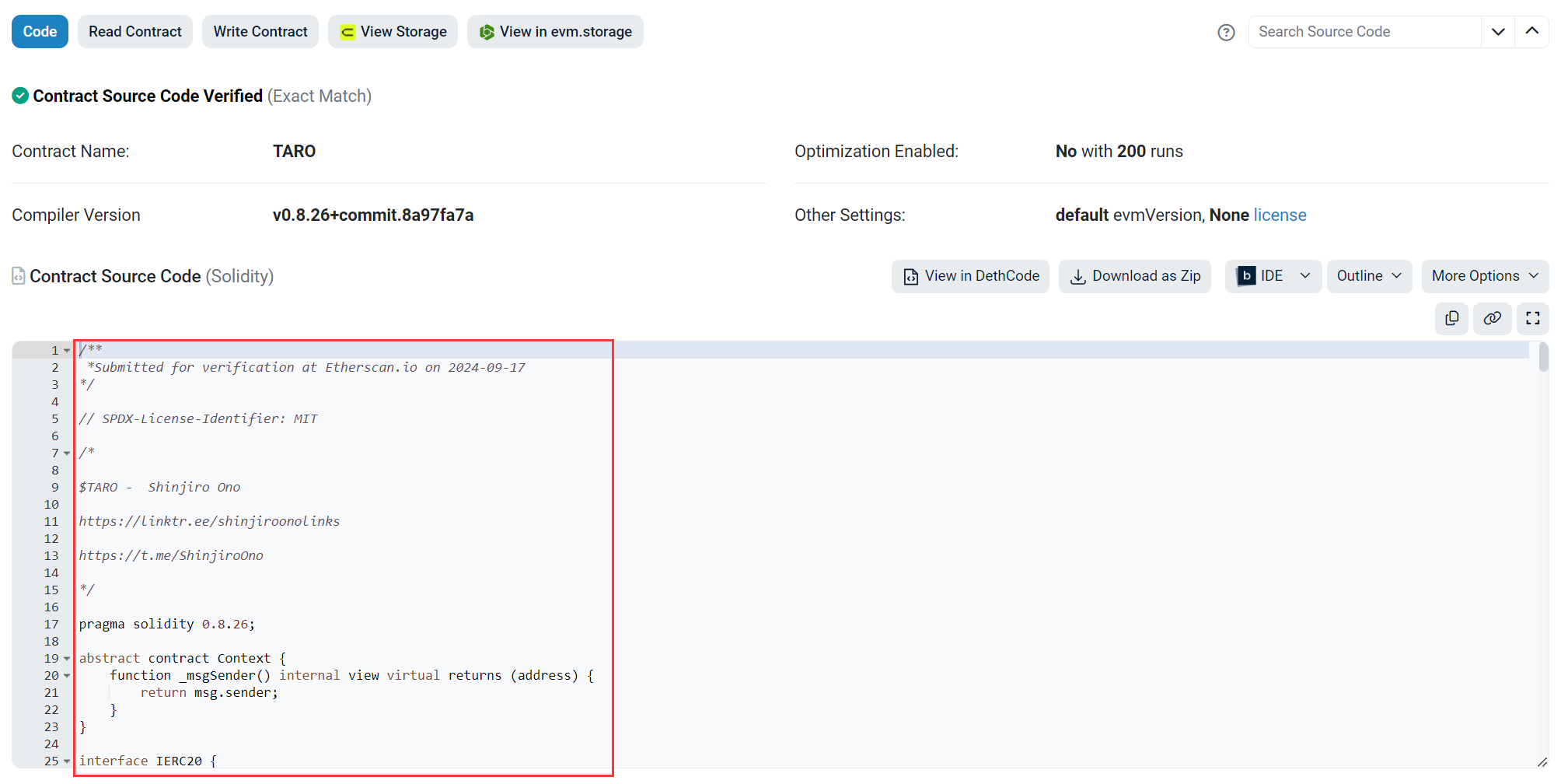}
  \caption{Source code of Etherscan smart contract}
  \label{fig:eth}
\end{figure}

As shown in Fig. \ref{fig:eth}, some smart contract vulnerability detection efforts are implemented based on source code \cite{Feist_2019_Slither, brent2020ethainter, schneidewind2020ethor}. While smart contracts deployed on blockchain systems are publicly transparent, it is not mandatory for contract developers to publish the source code. As a result, less than four of smart contracts on Ether are open source \cite{Garamvolgyi_2022_Utilizing}. Although some studies have been conducted to implement smart contract vulnerability detection from a bytecode perspective \cite{brent2018vandal, Veloso__Conkas, 10.1145/3543507.3583367}, a simple piece of bytecode or opcode is difficult to provide explicit vulnerability characterization, limiting its effectiveness in smart contract vulnerability detection \cite{tsankov2018securify}. To overcome this limitation, this paper extracts CFG from smart contract bytecode, which contains rich smart contract business logic and thus helps to realize BERT-based smart contract detection work more effectively.

In order to be able to more effectively detect vulnerabilities in bytecode-level smart contracts, this paper designs SmartBugBert, an efficient smart contract vulnerability detection method based on the BERT extension. To further improve the detection effect of smart contract vulnerabilities, this paper also combines the control flow graph in the static analysis technique and integrates the multi-dimensional detection method. The method is capable of detecting vulnerabilities from the smart contract bytecode and identifies four vulnerabilities: reentry vulnerability, arithmetic vulnerability, self-destructing contract, and timestamp dependency vulnerability \cite{Zhuang_2020_Smart, luo2024scvhunter, zhen2024gnn}.


\section{Background}
\label{sec:background}

This section provides essential background information on smart contract vulnerabilities, control flow graphs (CFGs), and BERT-based detection approaches to establish the foundation for our proposed bytecode-level vulnerability detection method.

\subsection{Smart Contract Vulnerabilities}

Smart contracts are self-executing digital agreements written in code that automatically enforce and execute predefined terms when specific conditions are met \cite{Ethereum}. These contracts operate on blockchain platforms and facilitate decentralized applications (DApps) and decentralized finance (DeFi) protocols \cite{li2024defitail, li2024stateguard}. However, due to their immutable nature, smart contract vulnerabilities can lead to significant financial losses if exploited \cite{li2024guardians, Zheng_2023_Blockchain-Based, slowmist}.

In this work, we focus on detecting four critical vulnerability types that frequently affect smart contracts:

\begin{itemize}
    \item \textbf{Transaction-Ordering Vulnerability (TOV)}: This vulnerability arises when the execution result of a transaction depends on the order in which transactions are mined, allowing attackers to manipulate transaction execution sequences for profit \cite{li2024cobra, Zhuang_2020_Smart}.
    
    \item \textbf{Access Control Vulnerability (ACV)}: This occurs when sensitive contract functions lack proper authorization checks, potentially allowing unauthorized users to execute privileged operations \cite{li2024guardians, Ghaleb_2023_AChecker}.
    
    \item \textbf{Self-Destruct Vulnerability (SDV)}: This vulnerability enables attackers to trigger a contract's self-destruct mechanism inappropriately, which can lead to permanent deletion of the contract and its assets \cite{li2020characterizing, Tikhomirov_2018_SmartCheck, torres2019art}.
    
    \item \textbf{Timestamp Dependency Vulnerability (TDV)}: This vulnerability exists when contracts rely on block timestamps for critical operations, which miners can manipulate within certain bounds \cite{li2024cobra, Nikolic_2018_Finding, Luu_2016_Making}.
\end{itemize}

The prevalence of these vulnerabilities in smart contracts, especially in emerging ecosystems like NFT marketplaces \cite{kong2024characterizing, niu2024unveiling, ma2025uncovering}, highlights the urgent need for effective detection techniques \cite{Yang_2023_Definition, Chen_2020_Survey}.

\subsection{Control Flow Graphs for Smart Contracts}

A CFG is a representation of all paths that might be traversed through a program during its execution \cite{Mossberg_2019_Manticore, wang2023smart}. In the context of smart contracts, CFGs provide valuable structural information about the contract's execution flow and help identify potential vulnerability patterns \cite{li2024cobra}.

Traditional approaches to vulnerability detection often rely solely on bytecode information without considering the control flow structure, which can lead to false positives or missed vulnerabilities \cite{Feist_2019_Slither, DurieuxEtAl2020ICSE}. Our approach addresses this limitation by recovering the CFG from the contract bytecode and extracting vulnerable CFG fragments that contain potential vulnerabilities. This approach enables more targeted and efficient vulnerability detection compared to analyzing the entire contract code \cite{mao2024scla, Xu_2023_SoK}.

\subsection{BERT-based Smart Contract Analysis}

Recent advancements in natural language processing, particularly the Bidirectional Encoder Representations from Transformers (BERT) model, have shown promising results in code analysis tasks \cite{li2023overview, Wang_2022_Unified}. BERT's ability to capture contextual relationships in sequences makes it well-suited for analyzing program code and identifying complex patterns \cite{Asudani_2023_Impact}.

For smart contract vulnerability detection, BERT can be fine-tuned to extract features from the CFG that represent potential vulnerable patterns \cite{li2024cobra, Samreen_2021_SmartScan, 10646885}. This approach offers advantages over traditional feature engineering methods as it can automatically learn relevant features from the data \cite{tann2019safer}.

Our work builds upon these foundations by combining BERT-based feature extraction with statistical semantic features and using LightGBM for classification. This integrated approach allows for more comprehensive vulnerability detection that considers both the semantic context and the control flow structure of smart contracts \cite{DurieuxEtAl2020ICSE, diAngeloEtAl2023ASE, chen2024improving, shang2025cegt, 10700860}.

\section{Method}
\begin{figure}[h]
  \centering
  \includegraphics[width=\linewidth]{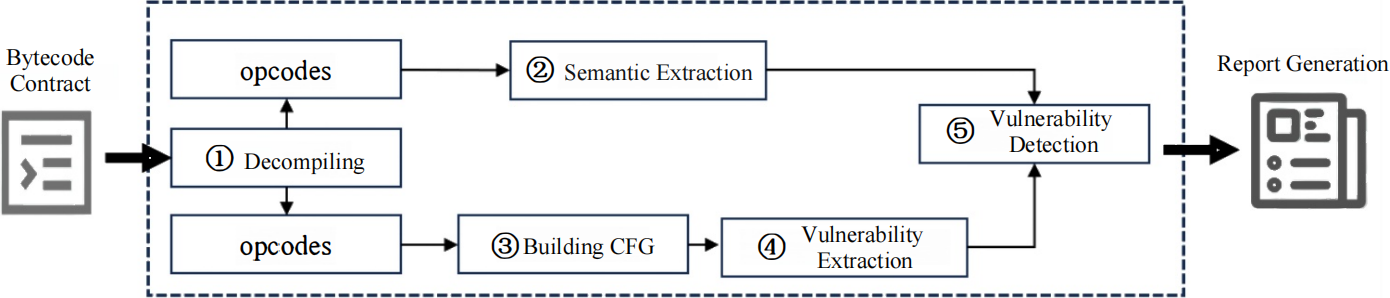}
  \caption{SmartBugBert Framework}
  \label{fig:smbert}
\end{figure}

In this section, SmartBugBERT smart contract vulnerability detection method is proposed, which mainly consists of three major parts: semantic extraction module, bytecode-level CFG module construction, and CFG vulnerability fragment extraction module. Specifically, the implementation steps of SmartBugBERT are shown in Fig. \ref{fig:smbert}: The context information, i.e., opcode information, is extracted from the collected bytecode-level smart contracts using  (1) decompilation module \cite{li2021hybrid}. After filtering the opcodes with the same function in the opcode information and (2) extracting the semantics; from the sequence of opcodes obtained from the decompiler module, (3) construct the control flow graph of the smart contract and extract the CFG fragments with vulnerabilities through (4) vulnerability fragments, utilize the Bert feature extraction and fuse it with the semantic features; the fused full features are sent to the classification module to complete the (5) vulnerability detection task and generate the report.

\subsection{Semantic Extraction Module}
Bytecode is stored on the blockchain as a string of hexadecimal numbers. Unlike source code, bytecode is completely open and transparent and can be easily accessed from each contract \cite{li2017discovering}. The bytecode-based semantic extraction module has two major steps: bytecode decompilation and feature extraction.

\begin{figure}[h]
  \centering
  \includegraphics[width=0.6\linewidth]{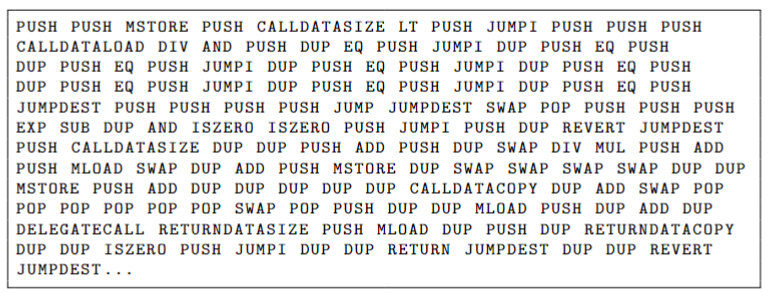}
  \caption{Optimized Opcode Sequence}
  \label{fig:byte}
\end{figure}

First, consider that smart contract bytecode is not easy to read for humans and does not have any semantic information. In this paper, we convert the bytecode into equivalent opcodes that are easy to be understood by humans through the pyevmasm disassembler to facilitate semantic extraction of the contract \cite{li2024detecting}. As shown in Fig. \ref{fig:byte}, the initial sequence of opcodes is redundant, and for better semantic extraction, SmartBugBERT optimizes the representation of operands with the same behavior, e.g., DUP1 and DUP2 are both considered as DUP; PUSH1 and PUSH2 are both considered as PUSH \cite{li2025scalm}.

Second, the optimized opcodes are statistically measured using TF-IDF, which describes the importance of a given opcode in a certain vulnerability category.TF-IDF measures the importance of a single opcode by the product of two parameters, the word frequency (TF) and the inverse document frequency (IDF).TF reflects the frequency of occurrence of a word in a document, while the IDF describes the rarity of a single opcode in the entire rarity of a single operand in the entire document collection. In general, the closer the IDF value is to 0, the more common the word is, and conversely, the more representative the word corresponding to the operand is \cite{wang2024smart}.

For example, among them PUSH, DUP, SWAP and POP are the four most commonly used in smart contracts. These opcodes are all related to stack operations. Since EVMs are stack-based, almost any operation, such as defining variables and functions, performing arithmetic operations (pressing data into the EVM stack), swapping elements, and deleting variables, requires stack operations. As a result, the IDFs of these opcodes are almost semantically unimportant. However, the IDF value of the opcode SELFDESTRUCT (byte value 0xFF) in a contract with a self-destruct vulnerability will exceed the IDF value of a contract with other vulnerabilities \cite{10707457}. Therefore, the statistical characterization of opcodes can be used to detect vulnerabilities in contracts, and it reflects the characterization of contract vulnerabilities to some extent from an EVM perspective.

\subsection{Control Flow Diagram Building Blocks at the Byte-Code Level}
The structure of a CFG is represented as $(N, E, N_{entry}, N_{exit})$. Here, $N$ denotes the set of nodes, where each node represents a sequence of instructions executed sequentially, called a Basic Block (BB). $E(B_i, B_j)$ represents the set of directed edges, indicating the jump relationship between basic blocks, where the control flow jumps from block $B_i$ to $B_j$.  

The jumps between basic blocks are implemented using the JUMP and JUMPI operations, with the jump target starting at a JUMPDEST. For each basic block, the entry point $N_{entry}$ and the exit point $N_{exit}$ are unique. This allows for information propagation between different blocks.

The general process of smart contract bytecode CFG generation is as follows: through the disassembly operation is converted into operation code (Opcode), and then through the division of each independent Basic Block (Basic Block), and finally for each Basic Block to add the jump relationship to get the final result of the CFG. among them, the determination of the Basic Block is crucial, in order to better determine the Basic Block, this paper sets the following rules:
\begin{itemize}
\item  The first instruction (opcode PUSH) of the decompiled EVM instruction sequence is the start instruction of the basic block;
\item  When the JUMPDEST opcode is encountered, it locates the start instruction of the target basic block, marking the entrance to the basic block;
\item  When the operation codes JUMP, JUMPI, STOP, RETURN, INVALID, REVERT, SELFDESTRUCT, and SUICIDE are encountered, they represent the end of the basic block;
\item  The sequence of instructions between the start and end instructions constitutes a complete basic block.
\end{itemize}

According to the above rules, all basic blocks can be divided, and the specific implementation of the basic block division algorithm is shown in Algorithm \ref{alg:CFG_construction}. When the start instruction is encountered, a new basic block is created and the current block is added to the list of basic blocks ; when the end instruction is encountered, the end of the current basic block is marked and added to the list of basic blocks. after the traversal of the EVM instruction sequence is finished, the last basic block is added to the list, completing the process of dividing basic blocks.

\begin{algorithm}
\caption{Control Flow Graph Construction Algorithm}
\label{alg:CFG_construction}
\begin{algorithmic}[1]
\Require Smart Contract Bytecode
\Ensure Control Flow Graph (CFG)
\State \textbf{Input:} \textit{bytecode}
\State \textbf{Output:} \textit{cfg}
\State
\State // Initialize basic block dictionary
\State \textit{basic\_block\_list} $\gets$ \{\}
\State // Initialize instruction dictionary
\State \textit{\_instructions\_dict} $\gets$ \{\}
\State // Initialize basic block
\State \textit{bb} $\gets$ \{\}
\State // Decompile bytecode into EVM opcode sequence
\State \textit{OpcodeSeq} $\gets$ \textsc{Decompile}(\textit{bytecode})
\For{each \textit{op} in \textit{OpcodeSeq}}
    \State Add (\texttt{pyevmasm.disassemble\_all.pc}, \textit{op}) to \textit{\_instructions\_dict}
    \If{\textit{op} is \texttt{JUMPDEST}}
        \State // Set the previous instruction as the end of \textit{bb}
        \State Set end instruction of \textit{bb} to previous \textit{op}
        \State Append \textit{bb} to \textit{basic\_block\_list}
        \State // Create a new basic block
        \State \textit{bb} $\gets$ \{\}
        \State // Set \textit{op} as the start of the new basic block
        \State Set start instruction of \textit{bb} to \textit{op}
    \ElsIf{\textit{op} is \texttt{STOP}, \texttt{SELFDESTRUCT}, \texttt{RETURN}, \texttt{REVERT}, \texttt{INVALID}, \texttt{SUICIDE}, \texttt{JUMP}, or \texttt{JUMPI}}
        \State // Set current instruction as the end of \textit{bb}
        \State Set end instruction of \textit{bb} to \textit{op}
        \State Append \textit{bb} to \textit{basic\_block\_list}
        \State // Create a new basic block
        \State \textit{new\_bb} $\gets$ \{\}
        \State // The next instruction becomes the start of the new basic block
        \State Set start instruction of \textit{new\_bb} to next instruction
    \EndIf
\EndFor
\For{each \textit{bb} in \textit{basic\_block\_list}}
    \State \textsc{AddEdge}(\textit{bb})
\EndFor
\State \Return \textit{cfg}
\end{algorithmic}
\end{algorithm}

\begin{figure}[h]
  \centering
  \includegraphics[width=0.8\linewidth]{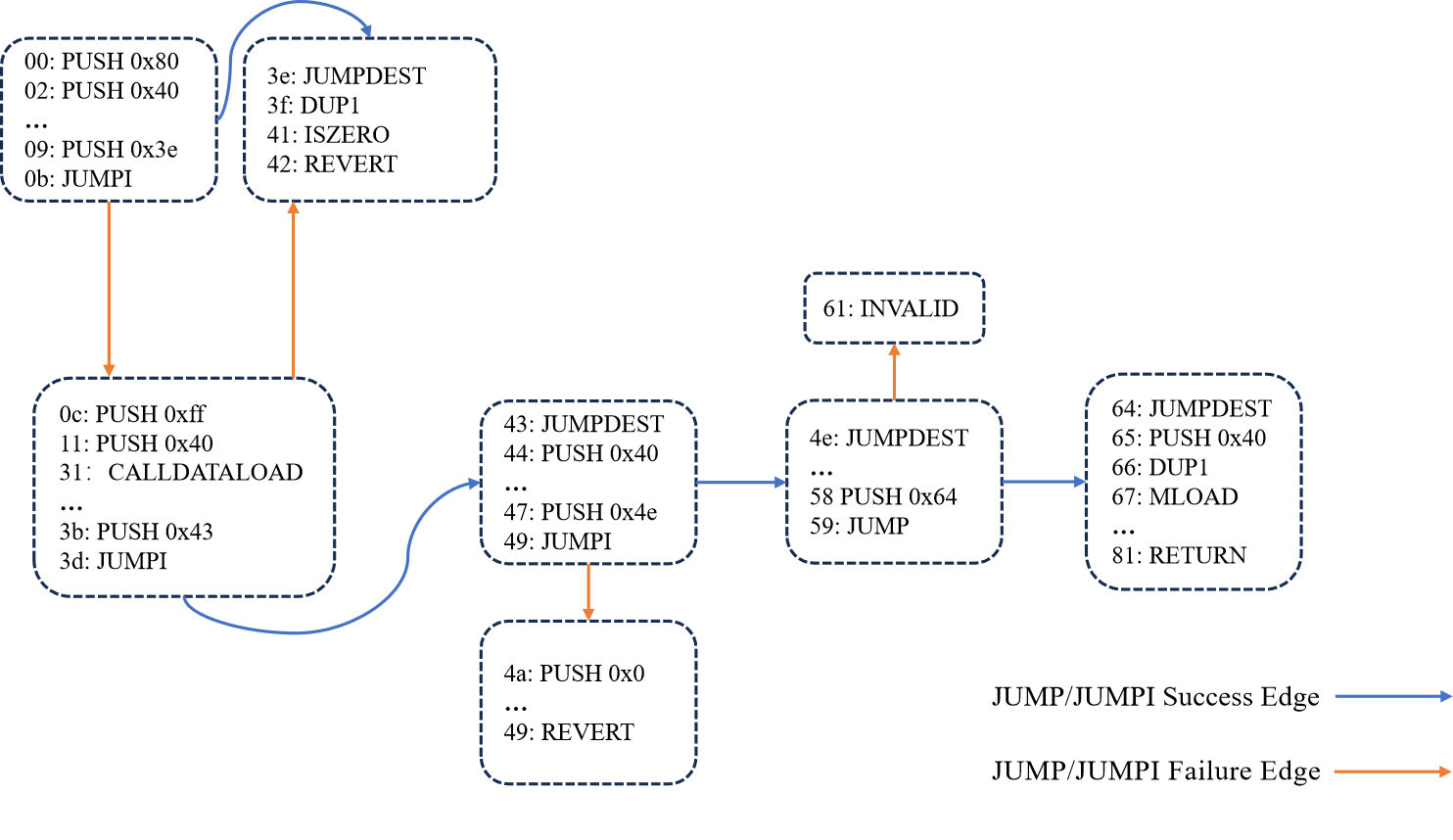}
  \caption{Control Flow Graph at the Bytecode Level}
  \label{fig:CFG}
\end{figure}

At the end of basic block generation, the edges of the control flow graph are generated. The next jump position of each basic block is obtained by traversing the basic blocks, and the original basic block and the basic block where the next jump is located form a directed edge, which is regarded as the directed edge of CFG. Basic block jump is divided into conditional jump (JUMPI) and unconditional jump (JUMP). JUMP directly from the top of the stack to read the address to jump.JUMPI each time from the stack to read two pieces of data, the first piece of data as the destination address, the second piece of data as a judgment condition, if the judgment condition is valid, the algorithm jumps to the destination address; Otherwise, the jump to the basic block JUMPI address the next block. Specifically as shown in Fig \ref{fig:CFG}.

\subsection{Control Flow Graph Vulnerability Fragment Extraction Module}
\begin{figure}[h]
  \centering
  \includegraphics[width=0.6\linewidth]{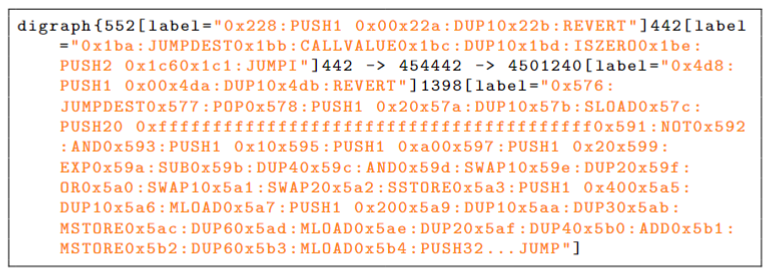}
  \caption{Control Flow Graph}
  \label{fig:BYTE2}
\end{figure}

As shown in Figure \ref{fig:BYTE2}, SmartBugBERT outputs the constructed bytecode-level CFG via a .dot file, with each basic block wrapped by "[ ]". Currently, most machine learning methods are difficult to process long text effectively. Meanwhile, in the process of annotating blank smart contract datasets, it is found that vulnerabilities tend to occur only in relation to a single or its associated function. Obviously, filtering irrelevant CFGs and retaining CFGs with vulnerability parts are only effective for vulnerability detection.

In order to realize the extraction of vulnerability fragments, screening is carried out from the following aspects: 

(1) The specific function in which the vulnerability occurs is first identified during the data labeling process. After labeling, the vulnerability location of the function is accurately identified by mapping the first four bytes of the function signature (the first four bytes of keccak256(functionSignature)) as a function selector) to the corresponding basic block in the control flow graph. 

(2) For contracts without published source code, pattern matching is utilized to match to fragments with vulnerabilities. For example, self-destruct vulnerabilities and timestamp dependency vulnerabilities match basic blocks with SELFDESTRUCT and TIMESTAMP opcodes. Reentry vulnerabilities match basic blocks for the presence of external contract calls (e.g., CALL, DELEGATECALL) that subsequently modify state (e.g., SSTORE) and associated jump basic blocks. Arithmetic vulnerabilities, on the other hand, match basic blocks where the arithmetic opcodes ADD, SUB, MUL, and DIV are present. Finally, the extracted vulnerability fragments are fed into the BERT model for feature extraction to obtain high-dimensional semantic representations. These representations capture the deep structure and contextual information of the vulnerability fragments and help in further vulnerability classification and detection.

\begin{figure}[h]
  \centering
  \includegraphics[width=\linewidth]{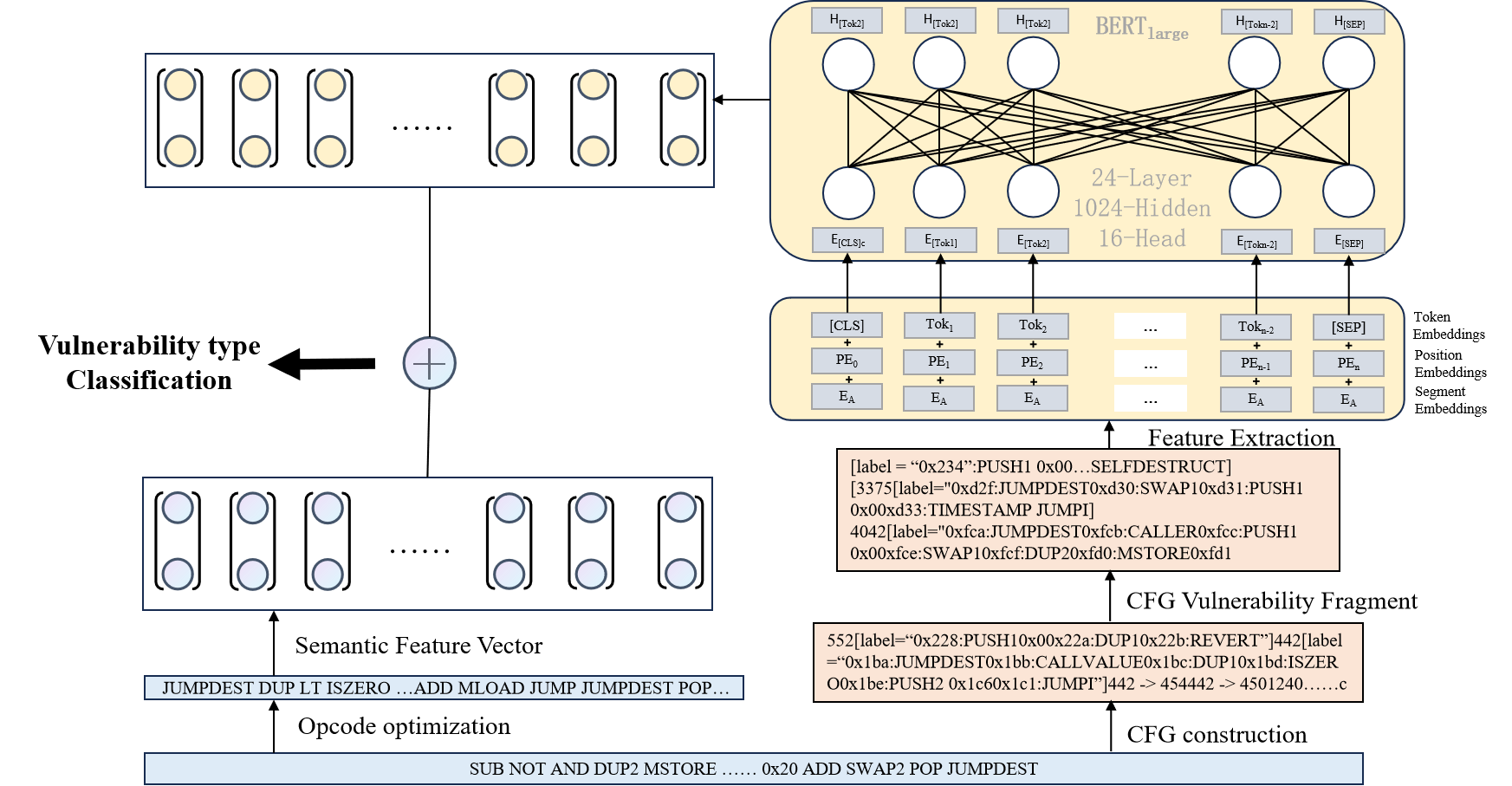}
  \caption{SmartBugBert Smart Contract Vulnerability Detection}
  \label{fig:SmartBugBert}
\end{figure}

\subsection{Model structure}
As in Fig \ref{fig:SmartBugBert}, the opcode sequence obtained from decompiling the original bytecode is processed through semantic extraction and CFG vulnerability fragment extraction. The opcode sequence OP = {OP1, OP2, ..., OPn} contains n opcodes. This OP is subjected to the optimization in Section 4.2.1 to obtain a relatively pure opcode sequence PureOP = {P1, P2, ..., P3}, and the number of opcode types after the statistical optimization is 80. Then, the PureOP sequence is converted to an 80-dimensional feature representation feature1 using TF-IDF to facilitate the training of the model.

Simple dependency opcodes can show the semantic features of self-destructing contracts and timestamp dependency vulnerabilities more clearly, however, for reentry vulnerabilities or arithmetic vulnerabilities that involve complex logic execution such as external calls and arithmetic processing, their semantic features are relatively vague and difficult to recognize. Therefore, SmartBugBERT recovers CFGs of bytecode-level smart contracts to show the complex execution logic. Usually, smart contract feature extraction mostly adopts Word2Vec model unidirectional or shallow contextual understanding to construct word vectors, which cannot adequately capture the bidirectional contextual information of words. In actual semantics, constructing word vectors based on unidirectional or shallow contextual understanding cannot fully capture the bidirectional contextual information of words. In arithmetic vulnerability detection, it is common to focus on opcodes related to addition, multiplication, subtraction and division, i.e., instructions such as ADD, MUL, SUB and DIV. However, if during the execution of these arithmetic instructions, comparison opcodes such as LT, GT, EQ, etc., are present in conjunction with JUMPI (conditional jumps) for exception handling, the logic can be considered to have included an overflow check. Therefore, in this case, it should not be concluded that there is an arithmetic vulnerability because the program has effectively prevented the risk of overflow through the conditional judgment and exception handling mechanism. Therefore, I use the BERT model to capture the rich contextual information in CFG.

To address the input length limitation of BERT models, this paper extracts vulnerable CFG fragments through the CFG Vulnerability Snippet Extraction module, which are then fed into BERT to obtain representations of contract vulnerability logic.

BERT transforms each opcode $\text{OP}_i$ in the sequence into word embeddings, position embeddings, and segment embeddings. These vectors are combined through addition to form a comprehensive composite embedding feature vector $\text{Feature}_2$ (eq. \ref{eq:embedding}):

\begin{equation}
\text{Feature}_2(x_i) = W_{\text{token}}(x_i) + W_{\text{position}}(i) + W_{\text{segment}}(s_i)
\label{eq:embedding}
\end{equation}

After the embedding layer, BERT employs a 24-layer Transformer encoder for feature extraction. Each Transformer encoder layer utilizes the self-attention mechanism to model contextual relationships in the input sequence. The self-attention mechanism is formulated as (eq. \ref{eq:attention}):

\begin{equation}
\text{Attention}(Q, K, V) = \text{softmax}\left(\frac{QK^T}{\sqrt{d_k}}\right)V
\label{eq:attention}
\end{equation}

The self-attention mechanism determines how much information to extract from the value vectors $V$ by computing the similarity between query vectors $Q$ and key vectors $K$. After processing through the Transformer layers, the final feature representation is obtained from the hidden states of the last layer: $H = (h_1, h_2, \dots, h_n)$, where $h_i$ is the hidden state vector of the $i$-th token in the input sequence. These hidden state vectors can be regarded as deep feature representations extracted by BERT.

Subsequently, the semantic features are fused with $\text{Feature}_2$ as input for downstream tasks. As shown in Figure 4.7, LightGBM is adopted as the classifier for contract vulnerability detection. LightGBM is an improved model based on GBDT, employing a leaf-wise strategy to control model complexity. As shown in Equation \eqref{eq:loss}, the objective function is enhanced with second-order Taylor expansion and regularization terms (eq. \ref{eq:loss}):

\begin{equation}
L_n = \sum_{i=1}^n l(y^i, \hat{y}^{i}_{n-1} + f_n(x^i)) + \delta T + \frac{1}{2}\sigma\sum_{j=1}^T \omega_j^2
\label{eq:loss}
\end{equation}

where $x^i$ denotes the $i$-th sample, $y^i$ represents its corresponding label, $l$ is the original loss function, $L_n$ indicates the regularized objective function at the $n$-th iteration, $f_n$ is the model at the $n$-th iteration, $\delta$ and $\sigma$ are parameters, $T$ is the number of leaf nodes, and $\omega_j$ is the output value of the $j$-th leaf node.

\section{Performance Analysis}
This section analyzes smart contract security at the bytecode level, systematically evaluates the capability of the SmartBugBERT model in detecting four types of vulnerabilities (RA, AV, SD, and TDV), and reports the Precision, Recall, and F1-Score on the test dataset.

\subsection{Experimental Data}
Since current research does not provide annotated datasets for bytecode-level smart contract vulnerability detection, and to ensure the authenticity of SmartBugBERT's effectiveness, real smart contracts were collected from Ethersca and annotated using existing smart contract detection tools \cite{10486822, LI2026122645}. First, Google BigQuery was used to collect 14,289 smart contract addresses. Then, Python scripts were employed to request the Solidity source code files and bytecode files of these smart contracts from Etherscan, establishing corresponding relationships. After filtering out duplicate contracts and those without source code, 9,346 unique bytecode contract files were obtained. Finally, existing contract vulnerability detection tools were used to annotate these contracts. Due to differences in the capabilities of various tools, we selected more advanced tools to collect contracts with RA, AV, SD, and TDV vulnerabilities as accurately as possible. Oyente was used to identify RA and TDV vulnerabilities \cite{Luu_2016_Making}, MAIAN to identify SD vulnerabilities, and Osiris to identify AV vulnerabilities \cite{torres2018osiris}. After filtering out contracts without detected vulnerabilities, 6,157 processed contracts and their corresponding vulnerability labels were collected.

\subsection{Experimental Environment}
The model's training and prediction processes were conducted on a server with the following hardware configuration: a Xeon(R) Platinum 8362 CPU, 60GB RAM, and a GeForce RTX 3090 GPU. The operating system was Ubuntu 20.04, running Python 3.8 and PyTorch.

\subsection{Evaluation Metrics}
To accurately and reasonably evaluate the performance of the SmartBugBERT model, we selected Precision, Recall, and F1-Score as evaluation metrics. These metrics are calculated using Equations~\eqref{eq:precision}, \eqref{eq:recall}, and \eqref{eq:f1}. Here, True Positive (TP) represents the number of smart contracts where vulnerabilities were correctly detected. False Negative (FN) represents the number of smart contracts that actually contain vulnerabilities but were not correctly identified by the model. False Positive (FP) represents the number of smart contracts incorrectly flagged as vulnerable by the model when they were not. True Negative (TN) represents the number of smart contracts correctly identified as non-vulnerable.

\textbf{Precision (PRE)}: The proportion of smart contracts correctly identified as vulnerable among all contracts flagged as vulnerable by the model.
\begin{equation}
    \text{Precision}: PRE = \frac{TP}{TP + FP} \label{eq:precision}
\end{equation}

\textbf{Recall (REC)}: The proportion of vulnerable smart contracts correctly identified by the model among all actually vulnerable contracts.
\begin{equation}
    \text{Recall}: REC = \frac{TP}{TP + FN} \label{eq:recall}
\end{equation}

\textbf{F1-Score (F1)}: The harmonic mean of Precision and Recall, used to comprehensively represent their performance.
\begin{equation}
    \text{F1-Score}: F1 = 2 \cdot \frac{PRE \cdot REC}{PRE + REC} \label{eq:f1}
\end{equation}

\subsection{Experimental Results and Analysis}
To demonstrate the effectiveness of SmartBugBert in detecting bytecode-level smart contract vulnerabilities, the dataset was divided into an 80\% training set and a 20\% test set. After training SmartBugBert on the training set, its performance was evaluated on the test set.

To further evaluate the contract vulnerability detection effectiveness of our method, we compared it with two other smart contract vulnerability detection approaches: SaferSC and Oyente \cite{Luu_2016_Making, grieco2020echidna, wustholz2020harvey}. As shown in Table~\ref{tab:comparison}, our method achieved significant improvements in vulnerability detection compared to both SaferSC and Oyente. Specifically, our method improved precision by 41.19\% over SaferSC and 48.86\% over Oyente, while improving recall by 39.51\% over SaferSC and 45.25\% over Oyente. The relatively poor performance of Oyente is attributed to its inability to detect self-destructing contracts, which reflects the limitations of traditional contract vulnerability detection methods.

\begin{table}[h]
\centering
\caption{Comparative experiments of different methods}
\label{tab:comparison}
\begin{tabular}{lccc}
\hline
Method & PRE & REC & F1-Score \\
\hline
SaferSC & 49.43\% & 52.25\% & 50.80\% \\
Oyente & 41.76\% & 46.51\% & 44.00\% \\
SmartBugBert & 90.62\% & 91.76\% & 91.19\% \\
\hline
\end{tabular}
\end{table}

\subsection{Ablation Study}
To demonstrate the impact of CFG information on contract vulnerability detection, we designed three ablation experiments: (1) using only opcode semantic features, (2) using only CFG features, and (3) combining semantic features with CFG features to form full features. The experimental results are shown in Table~\ref{tab:features}.

\begin{table}[h]
\centering
\caption{Comparison results of different features for contract vulnerability detection}
\label{tab:features}
\begin{tabular}{lccc}
\hline
Feature Selection & PRE & REC & F1-Score \\
\hline
Semantic features only & 66.52\% & 67.61\% & 67.26\% \\
CFG features only & 83.27\% & 87.41\% & 86.76\% \\
Full features & 90.62\% & 91.76\% & 91.19\% \\
\hline
\end{tabular}
\end{table}

From Table~\ref{tab:features}, we observe that relying solely on semantic features (optimized opcode sequences) for guiding the model to detect smart contract vulnerabilities yields mediocre results, with an F1-Score of only 67.26\%. The precision reaches only 66.52\%. Additionally, as shown in Table~\ref{tab:vulnerability_types}, using only semantic features results in lower detection effectiveness for reentrancy vulnerabilities (RV) compared to other vulnerability types by approximately 5\%-19\%. This occurs because reentrancy vulnerabilities involve complex contract call logic that cannot be adequately expressed through opcode sequences alone. Therefore, we conclude that single opcode sequence features cannot accurately accomplish contract vulnerability detection tasks.


\begin{table}[h]
\centering
\caption{Comparison of precision results for different vulnerability types}
\label{tab:vulnerability_types}
\begin{tabular}{cccc}  
\hline
\multirow{2}{*}{Vulnerability Type} & \multicolumn{3}{c}{Precision} \\  
\cline{2-4}  
 & Semantic Only & CFG Only & Full Feature \\  
\hline
RV & 55.24\% & 75.67\% & 87.96\% \\
AV & 59.00\% & 80.57\% & 90.91\% \\
SD & 74.04\% & 90.87\% & 92.11\% \\
TDV & 66.67\% & 88.19\% & 89.89\% \\
\hline
\end{tabular}
\end{table}

As shown in Table~\ref{tab:vulnerability_types}, using CFG features improves detection for all vulnerability types to varying degrees. This occurs because CFG contains rich logical information about smart contract vulnerabilities, demonstrating that CFG information is effective for vulnerability detection tasks. After training the model on the combined semantic and CFG features (full features) on the training set and evaluating it on the test set, the results showed excellent performance: 90.62\% precision, 91.76\% recall, and 91.19\% F1-Score.

\begin{figure}[h]
\centering
\includegraphics[width=0.7\textwidth]{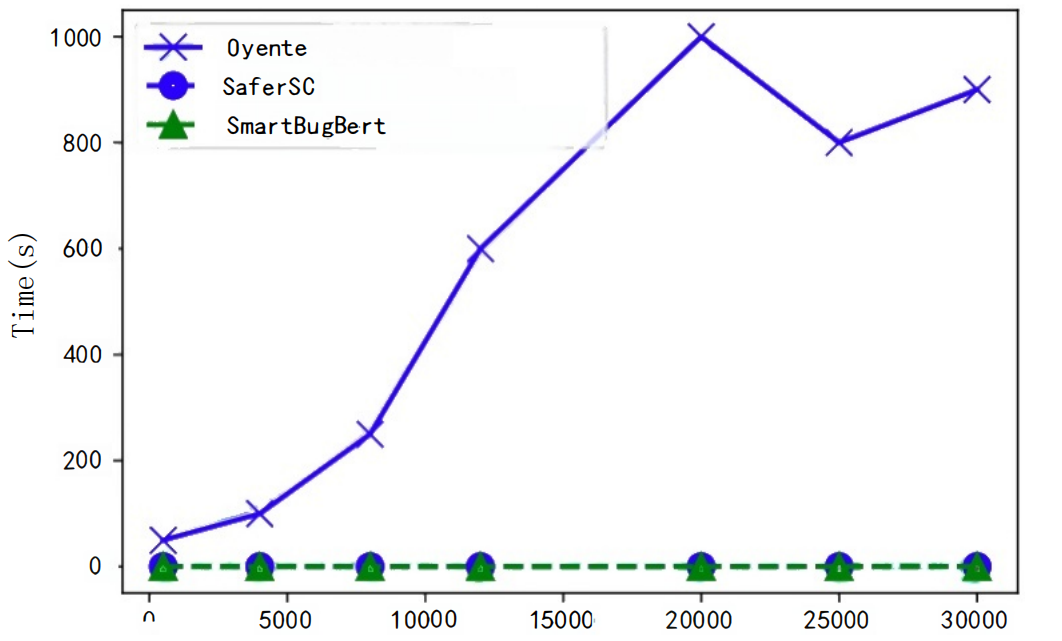}
\caption{Comparison of analysis time between symbolic execution and machine learning methods}
\label{fig:time_comparison}
\end{figure}

Figure~\ref{fig:time_comparison} shows that the vulnerability detection time of the symbolic tool Oyente increases significantly with the complexity of smart contracts (measured by opcode length). In contrast, the analysis time required by machine learning methods remains relatively stable. 

To compare the difference in analysis time between machine learning methods and Oyente, we calculated the average opcode length for different vulnerability categories and defined seven distinct code complexity levels. We then measured the average analysis time for ten smart contracts of each length and plotted the results to highlight the differences in time consumption between the two approaches.

\begin{table}[h]
\centering
\caption{Average execution time of vulnerability detection methods}
\label{tab:execution_time}
\begin{tabular}{lc}
\hline
Detection Method & Average Time (seconds) \\
\hline
Symbolic Execution (Oyente) & 528.57 \\
Machine Learning Method (SaferSC) & 0.23 \\
Machine Learning Method (SmartBugBERT) & 0.14 \\
\hline
\end{tabular}
\end{table}

Table~\ref{tab:execution_time} shows that the average vulnerability detection time for the symbolic execution tool is 528.57 seconds, while the proposed machine learning-based SaferSC and SmartBugBert methods have average detection times of 0.14 seconds and 0.23 seconds respectively. This difference arises because machine learning methods only detect specific vulnerabilities they were trained on, without performing comprehensive analysis of other characteristics of smart contracts \cite{mythril, Mossberg_2019_Manticore, chen2025chatgpt, mueller2018smashing, pakala}.

\section{Conclusion}

This paper introduces a BERT-based bytecode-level smart contract vulnerability detection method \cite{10.1145/3699597, song2025silence}. The approach first decompiles smart contract bytecode into opcode sequences and represents their semantic features through statistical characteristics. Unlike methods that solely use bytecode information, this chapter recovers the CFG from the bytecode level and extracts vulnerable CFG fragments containing potential vulnerabilities for efficient detection, while utilizing fine-tuned BERT to extract CFG features. Subsequently, the semantic features and CFG features are fused as input to the LightGBM classifier to accomplish the contract vulnerability detection task. Experimental results demonstrate that the proposed method can effectively detect Transaction-Ordering Vulnerability, Access Control Vulnerability, Self-Destruct Vulnerability, and Timestamp Dependency Vulnerability in contracts, achieving an excellent F1-Score of 91.19\% \cite{liu2025detecting, RuntimeVerification}. Regarding detection time, it also shows significant advantages compared to symbolic execution-based tools.

\bibliographystyle{ACM-Reference-Format}
\bibliography{paper1}

@inproceedings{Zhuang_2020_Smart,
  title = {Smart {{Contract Vulnerability Detection}} Using {{Graph Neural Network}}},
  booktitle = {Proceedings of the {{29th International Joint Conference}} on {{Artificial Intelligence}} (IJCAI)},
  author = {Zhuang, Yuan and Liu, Zhenguang and Qian, Peng and Liu, Qi and Wang, Xiang and He, Qinming},
  year = {2020},
  pages = {3283--3290},
  
}

@article{Xia_2021_Tradea,
  title = {Trade or {{Trick}}? {{Detecting}} and {{Characterizing Scam Tokens}} on {{Uniswap Decentralized Exchange}}},
  author = {Xia, Pengcheng and Wang, Haoyu and Gao, Bingyu and Su, Weihang and Yu, Zhou and Luo, Xiapu and et al.},
  year = {2021},
  journal = {Proceedings of the ACM on Measurement and Analysis of Computing Systems (POMACS)},
  volume = {5},
  number = {3},
  pages = {1--26}
}

@article{Ethereum,
  title={Ethereum: A secure decentralised generalised transaction ledger},
  author={Wood, Gavin and others},
  journal={Ethereum project yellow paper},
  volume={151},
  number={2014},
  pages={1--32},
  year={2014}
}

@inproceedings{Garamvolgyi_2022_Utilizing,
  title = {Utilizing Parallelism in Smart Contracts on Decentralized Blockchains by Taming Application-Inherent Conflicts},
  booktitle = {Proceedings of the {{44th International Conference}} on {{Software Engineering}} (ICSE)},
  author = {Garamvölgyi, Péter and Liu, Yuxi and Zhou, Dong and Long, Fan and Wu, Ming},
  year = {2022},
  pages = {2315--2326}
}

@article{Chen_2020_Survey,
  title = {A {{Survey}} on {{Ethereum Systems Security}}: {{Vulnerabilities}}, {{Attacks}}, and {{Defenses}}},
  author = {Chen, Huashan and Pendleton, Marcus and Njilla, Laurent and Xu, Shouhuai},
  year = {2020},
  journal = {ACM Computing Surveys},
  volume = {53},
  number = {3},
  pages = {1--43}
}

@misc{RuntimeVerification,
	title = {Runtime Verification},
	note = {\url{https://runtimeverification.com/}},
	year = {2024}
}

@inproceedings{Ghaleb_2023_AChecker,
  title = {{{AChecker}}: {{Statically Detecting Smart Contract Access Control Vulnerabilities}}},
  booktitle = {Proceedings of the {45th {IEEE}}/{{ACM}} {{International Conference}} on {{Software Engineering}} (ICSE)},
  author = {Ghaleb, Asem and Rubin, Julia and Pattabiraman, Karthik},
  year = {2023},
  pages = {945--956}
}

@inproceedings{Wang_2022_Unified,
  title = {Unified Abstract Syntax Tree Representation Learning for Cross-Language Program Classification},
  booktitle = {Proceedings of the {{30th IEEE}}/{{ACM International Conference}} on {{Program Comprehension}} (ICPC)},
  author = {Wang, Kesu and Yan, Meng and Zhang, He and Hu, Haibo},
  year = {2022},
  pages = {390--400}
}

@article{Asudani_2023_Impact,
  title = {Impact of Word Embedding Models on Text Analytics in Deep Learning Environment: A Review},
  author = {Asudani, Deepak Suresh and Nagwani, Naresh Kumar and Singh, Pradeep},
  year = {2023},
  journal = {Artificial Intelligence Review},
  volume = {56},
  number = {9},
  pages = {10345--10425}
}

@misc{mythril,
    title       = {Mythril},
    note       = {\url{https://mythril-classic.readthedocs.io/}},
    year        = {2024}
}

@inproceedings{Luu_2016_Making,
  title = {Making {{Smart Contracts Smarter}}},
  booktitle = {Proceedings of the {{ACM SIGSAC Conference}} on {{Computer}} and {{Communications Security}} (CCS)},
  author = {Luu, Loi and Chu, Duc-Hiep and Olickel, Hrishi and Saxena, Prateek and Hobor, Aquinas},
  year = {2016},
  pages = {254--269}
}

@inproceedings{tsankov2018securify,
  title={Securify: Practical security analysis of smart contracts},
  author={Tsankov, Petar and Dan, Andrei and Drachsler-Cohen, Dana and Gervais, Arthur and Buenzli, Florian and Vechev, Martin},
  booktitle={Proceedings of the {ACM SIGSAC Conference} on {Computer and Communications Security} (CCS)},
  pages={67--82},
  year={2018}
}

@inproceedings{Mossberg_2019_Manticore,
  title = {Manticore: {{A User-Friendly Symbolic Execution Framework}} for {{Binaries}} and {{Smart Contracts}}},
  booktitle = {Proceedings of the {34th {{IEEE}}/{{ACM International Conference}} on {{Automated Software Engineering}}} (ASE)},
  author = {Mossberg, Mark and Manzano, Felipe and Hennenfent, Eric and Groce, Alex and Grieco, Gustavo and Feist, Josselin and et al.},
  year = {2019},
  pages = {1186--1189},
}

@misc{Veloso__Conkas,
    author = {Veloso, Nuno},
    title = {Conkas: {{A Modular}} and {{Static Analysis Tool}} for {{Ethereum Bytecode}}},
    year = {2023},
    note={\url{https://github.com/nveloso/conkas/}}

}

@inproceedings{Feist_2019_Slither,
  title = {Slither: {{A Static Analysis Framework}} for {{Smart Contracts}}},
  booktitle = {Proceedings of the {{2nd IEEE}}/{{ACM}} {{International Workshop}} on {{Emerging Trends}} in {{Software Engineering}} for {{Blockchain}} (WETSEB)},
  author = {Feist, Josselin and Grieco, Gustavo and Groce, Alex},
  year = {2019},
  pages = {8--15}
}

@inproceedings{Tikhomirov_2018_SmartCheck,
  title = {{{SmartCheck}}: Static Analysis of Ethereum Smart Contracts},
  booktitle = {Proceedings of the {{1st International Workshop}} on {{Emerging Trends}} in {{Software Engineering}} for {{Blockchain}} (WETSEB)},
  author = {Tikhomirov, Sergei and Voskresenskaya, Ekaterina and Ivanitskiy, Ivan and Takhaviev, Ramil and Marchenko, Evgeny and Alexandrov, Yaroslav},
  year = {2018},
  pages = {9--16}
}

@article{Xu_2023_SoK,
  title = {{{SoK}}: {{Decentralized Exchanges}} ({{DEX}}) with {{Automated Market Maker}} ({{AMM}}) {{Protocols}}},
  author = {Xu, Jiahua and Paruch, Krzysztof and Cousaert, Simon and Feng, Yebo},
  year = {2023},
  journal = {ACM Computing Surveys},
  volume = {55},
  number = {11},
  pages = {1--50}
}

@inproceedings{diAngeloEtAl2023ASE,
  title = {{SmartBugs} 2.0: An Execution Framework for Weakness Detection in {Ethereum} Smart Contracts},
  author={di Angelo, Monika and Durieux, Thomas and Ferreira, Jo{\~a}o F. and Salzer, Gernot},
  booktitle={Proceedings of the {38th IEEE/ACM International Conference} on {Automated Software Engineering} (ASE)},
  year={2023}
}

@inproceedings{DurieuxEtAl2020ICSE,
  title={Empirical Review of Automated Analysis Tools on 47,587 {Ethereum} Smart Contracts},
  author={Durieux, Thomas and Ferreira, Jo{\~a}o F. and Abreu, Rui and Cruz, Pedro},
  booktitle={Proceedings of the {42nd ACM/IEEE International conference} on {software engineering} (ICSE)},
  pages={530--541},
  year={2020}
}

@inproceedings{Samreen_2021_SmartScan,
  title = {{{SmartScan}}: {{An}} Approach to Detect {{Denial}} of {{Service Vulnerability}} in {{Ethereum Smart Contracts}}},
  booktitle = {Proceedings of the {{4th IEEE}}/{{ACM}} {{International Workshop}} on {{Emerging Trends}} in {{Software Engineering}} for {{Blockchain}} (WETSEB)},
  author = {Samreen, Noama Fatima and Alalfi, Manar H.},
  year = {2021},
  pages = {17--26}
}

@inproceedings{Nikolic_2018_Finding,
  title = {Finding {{The Greedy}}, {{Prodigal}}, and {{Suicidal Contracts}} at {{Scale}}},
  booktitle = {Proceedings of the {{34th Annual Computer Security Applications Conference}} (ACSAC)},
  author = {Nikolić, Ivica and Kolluri, Aashish and Sergey, Ilya and Saxena, Prateek and Hobor, Aquinas},
  year = {2018},
  pages = {653--663}
}

@article{Zheng_2023_Blockchain-Based,
  title = {Blockchain-{{Based Decentralized Application}}: {{A Survey}}},
  author = {Zheng, Peilin and Jiang, Zigui and Wu, Jiajing and Zheng, Zibin},
  year = {2023},
  journal = {IEEE Open Journal of the Computer Society},
  volume = {4},
  pages = {121--133},
}

@inproceedings{Yang_2023_Definition,
  title = {Definition and {{Detection}} of {{Defects}} in {{NFT Smart Contracts}}},
  booktitle = {Proceedings of the 32nd {{ACM SIGSOFT International Symposium}} on {{Software Testing}} and {{Analysis}} (ISSTA)},
  author = {Yang, Shuo and Chen, Jiachi and Zheng, Zibin},
  year = {2023},
  pages = {373--384}
}

@article{tann2019safer,
  title={Towards safer smart contracts: A sequence learning approach to detecting security threats},
  author={Tann, Wesley Joon-Wie and Han, Xing Jie and Gupta, Sourav Sen and Ong, Yew-Soon},
  journal={arXiv preprint arXiv:1811.06632},
  year={2018}
}

@misc{slowmist,
    author = {SlowMist Zone},
	year = {2024},
	title = {SlowMist Hacked},
    note = {\url{https://hacked.slowmist.io/}}
}

@inproceedings{li2024cobra,
  title={COBRA: Interaction-Aware Bytecode-Level Vulnerability Detector for Smart Contracts},
  author={Li, Wenkai and Li, Xiaoqi and Li, Zongwei and Zhang, Yuqing},
  booktitle={Proceedings of the 39th IEEE/ACM International Conference on Automated Software Engineering (ASE)},
  pages={1358--1369},
  year={2024}
}

@inproceedings{kong2024characterizing,
  title={Characterizing the Solana NFT Ecosystem},
  author={Kong, Dechao and Li, Xiaoqi and Li, Wenkai},
  booktitle={Companion Proceedings of the ACM on Web Conference (WWW)},
  pages={766--769},
  year={2024}
}

@inproceedings{niu2024unveiling,
  title={Unveiling Wash Trading in Popular NFT Markets},
  author={Niu, Yuanzheng and Li, Xiaoqi and Peng, Hongli and Li, Wenkai},
  booktitle={Companion Proceedings of the ACM on Web Conference (WWW)},
  pages={730--733},
  year={2024}
}

@article{li2023overview,
  title={An overview of AI and blockchain integration for privacy-preserving},
  author={Li, Zongwei and Kong, Dechao and Niu, Yuanzheng and Peng, Hongli and Li, Xiaoqi and Li, Wenkai},
  journal={arXiv preprint arXiv:2305.03928},
  year={2023}
}

@inproceedings{li2024stateguard,
  title={StateGuard: Detecting State Derailment Defects in Decentralized Exchange Smart Contract},
  author={Li, Zongwei and Li, Wenkai and Li, Xiaoqi and Zhang, Yuqing},
  booktitle={Companion Proceedings of the ACM on Web Conference (WWW)},
  pages={810--813},
  year={2024}
}

@inproceedings{li2024defitail,
  title={DeFiTail: DeFi Protocol Inspection through Cross-Contract Execution Analysis},
  author={Li, Wenkai and Li, Xiaoqi and Zhang, Yuqing and Li, Zongwei},
  booktitle={Companion Proceedings of the ACM on Web Conference (WWW)},
  pages={786--789},
  year={2024}
}

@article{mao2024scla,
  title={SCLA: Automated Smart Contract Summarization via LLMs and Control Flow Prompt},
  author={Mao, Yingjie and Li, Xiaoqi and Li, Wenkai and Wang, Xin and Xie, Lei},
  journal={arXiv preprint arXiv:2402.04863},
  year={2024}
}

@inproceedings{li2020characterizing,
  title={Characterizing erasable accounts in ethereum},
  author={Li, Xiaoqi and Chen, Ting and Luo, Xiapu and Yu, Jiangshan},
 booktitle={Proceedings of the 23rd International Conference on Information Security (ISC)},
  pages={352--371},
  year={2020}
}

@article{li2024guardians,
  title={Guardians of the ledger: Protecting decentralized exchanges from state derailment defects},
  author={Li, Zongwei and Li, Wenkai and Li, Xiaoqi and Zhang, Yuqing},
  journal={IEEE Transactions on Reliability},
  year={2024}
}

@phdthesis{li2021hybrid,
  author  = {Li, Xiaoqi},
  title   = {Hybrid analysis of smart contracts and malicious behaviors in ethereum},
  school  = {Hong Kong Polytechnic University},
  year    = {2021}
}

@incollection{li2017discovering,
  author    = {Li, Xiaoqi and Yu, Le and Luo, Xiapu},
  title     = {On Discovering Vulnerabilities in Android Applications},
  booktitle = {Mobile Security and Privacy},
  year      = {2017},
  pages     = {155--166}
}

@inproceedings{10707457,
  author={Liu, Zekai and Li, Xiaoqi and Peng, Hongli and Li, Wenkai},
  booktitle={2024 IEEE International Conference on Web Services (ICWS)}, 
  title={GasTrace: Detecting Sandwich Attack Malicious Accounts in Ethereum}, 
  year={2024},
  volume={},
  number={},
  pages={1409-1411}
}

@misc{wang2024smart,
  title         = {Smart Contracts in the Real World: A Statistical Exploration of External Data Dependencies},
  author        = {Wang, Yishun and Li, Xiaoqi and Ye, Shipeng and Xie, Lei and Ju Xing},
  year          = {2024},
  archivePrefix = {arXiv},
  eprint        = {2406.13253},
  journal       = {arXiv preprint}
}

@misc{li2025scalm,
  title         = {SCALM: Detecting Bad Practices in Smart Contracts Through LLMs},
  author        = {Li, Zongwei and Li, Xiaoqi and Li, Wenkai and others},
  year          = {2025},
  archivePrefix = {arXiv},
  eprint        = {2502.04347},
  journal       = {arXiv preprint}
}

@inproceedings{li2024detecting,
  author    = {Li, Wenkai and Liu, Zheng and Li, Xiaoqi and others},
  title     = {Detecting Malicious Accounts in Web3 through Transaction Graph},
  booktitle = {Proceedings of the 39th IEEE/ACM International Conference on Automated Software Engineering (ASE)},
  year      = {2024},
  pages     = {2482--2483}
}

@inproceedings{brent2020ethainter,
  title     = {Ethainter: a smart contract security analyzer for composite vulnerabilities},
  author    = {Brent, Lexi and Grech, Neville and Lagouvardos, Sifis and Scholz, Bernhard and Smaragdakis, Yannis},
  booktitle = {Proceedings of the ACM SIGPLAN Conference on Programming Language Design and Implementation (PLDI)},
  pages     = {454--469},
  year      = {2020}
}

@article{brent2018vandal,
  title   = {Vandal: A scalable security analysis framework for smart contracts},
  author  = {Brent, Lexi and Jurisevic, Anton and Kong, Michael and Liu, Eric and Gauthier, Francois and Gramoli, Vincent and Holz, Ralph and Scholz, Bernhard},
  journal = {arXiv preprint arXiv:1809.03981},
  year    = {2018}
}

@inproceedings{grieco2020echidna,
  title     = {Echidna: effective, usable, and fast fuzzing for smart contracts},
  author    = {Grieco, Gustavo and Song, Will and Cygan, Artur and Feist, Josselin and Groce, Alex},
  booktitle = {Proceedings of the ACM SIGSOFT international symposium on software testing and analysis (ISSTA)},
  pages     = {557--560},
  year      = {2020}
}

@inproceedings{wustholz2020harvey,
  title     = {Harvey: A greybox fuzzer for smart contracts},
  author    = {W{\"u}stholz, Valentin and Christakis, Maria},
  booktitle = {Proceedings of the ACM Joint Meeting on European Software Engineering Conference and Symposium on the Foundations of Software Engineering (FSE)},
  pages     = {1398--1409},
  year      = {2020}
}

@article{mueller2018smashing,
  title   = {Smashing ethereum smart contracts for fun and real profit},
  author  = {Mueller, Bernhard},
  journal = {HITB SECCONF Amsterdam},
  volume  = {9},
  number  = {54},
  pages   = {4--17},
  year    = {2018}
}

@inproceedings{luo2024scvhunter,
  title     = {Scvhunter: Smart contract vulnerability detection based on heterogeneous graph attention network},
  author    = {Luo, Feng and Luo, Ruijie and Chen, Ting and Qiao, Ao and He, Zheyuan and Song, Shuwei and Jiang, Yu and Li, Sixing},
  booktitle = {Proceedings of the IEEE/ACM 46th international conference on software engineering (ICSE)},
  pages     = {1--13},
  year      = {2024}
}

@inproceedings{schneidewind2020ethor,
  title     = {ethor: Practical and provably sound static analysis of ethereum smart contracts},
  author    = {Schneidewind, Clara and Grishchenko, Ilya and Scherer, Markus and Maffei, Matteo},
  booktitle = {Proceedings of the ACM SIGSAC Conference on Computer and Communications Security (CCS)},
  pages     = {621--640},
  year      = {2020}
}

@inproceedings{torres2019art,
  title     = {The art of the scam: Demystifying honeypots in ethereum smart contracts},
  author    = {Torres, Christof Ferreira and Steichen, Mathis and others},
  booktitle = {Proceedings of the 28th USENIX Security Symposium (USENIX Security)},
  pages     = {1591--1607},
  year      = {2019}
}

@inproceedings{torres2018osiris,
  title     = {Osiris: Hunting for integer bugs in ethereum smart contracts},
  author    = {Torres, Christof Ferreira and Sch{\"u}tte, Julian and State, Radu},
  booktitle = {Proceedings of the 34th annual computer security applications conference (ACSAC)},
  pages     = {664--676},
  year      = {2018}
}

@misc{pakala,
  title  = {Offensive vulnerability scanner for ethereum, and symbolic execution tool for the Ethereum Virtual Machine},
  author = {Palkeo},
  note   = {\url{https://github.com/palkeo/pakala/}},
  year   = {2025}
}

@article{chen2025chatgpt,
  title   = {When chatgpt meets smart contract vulnerability detection: How far are we?},
  author  = {Chen, Chong and Su, Jianzhong and Chen, Jiachi and Wang, Yanlin and Bi, Tingting and Yu, Jianxing and Wang, Yanli and Lin, Xingwei and Chen, Ting and Zheng, Zibin},
  journal = {ACM Transactions on Software Engineering and Methodology (TOSEM)},
  volume  = {34},
  number  = {4},
  pages   = {1--30},
  year    = {2025}
}

@inproceedings{chen2024improving,
  title     = {Improving smart contract security with contrastive learning-based vulnerability detection},
  author    = {Chen, Yizhou and Sun, Zeyu and Gong, Zhihao and Hao, Dan},
  booktitle = {Proceedings of the IEEE/ACM 46th International Conference on Software Engineering (ICSE)},
  pages     = {1--11},
  year      = {2024}
}

@article{liu2025detecting,
  title   = {Detecting Smart Contract State-Inconsistency Bugs via Flow Divergence and Multiplex Symbolic Execution},
  author  = {Liu, Yinxi and Meng, Wei and Zhang, Yinqian},
  journal = {Proceedings of the ACM on Software Engineering},
  volume  = {2},
  number  = {FSE},
  pages   = {22--43},
  year    = {2025}
}

@inproceedings{song2025silence,
  title     = {Silence False Alarms: Identifying Anti-Reentrancy Patterns on Ethereum to Refine Smart Contract Reentrancy Detection},
  author    = {Song, Qiyang and Huang, Heqing and Jia, Xiaoqi and Xie, Yuanbo and Cao, Jiahao},
  booktitle = {Proceedings of the Network and Distributed System Security (NDSS)},
  year      = {2025},
  pages     = {1--18}
}

@article{shang2025cegt,
  title   = {CEGT: Smart contract vulnerability detection via Connectivity-Enhanced GCN-Transformer},
  author  = {Shang, Jiandong and Li, Jiaru and Sui, Yizhe and Guo, Hengliang and Gao, Xu and Zhang, Dujuan and Guo, Yang and Wu, Gang},
  journal = {Journal of Systems and Software},
  pages   = {112454--112465},
  year    = {2025}
}

@article{zhen2024gnn,
  title   = {DA-GNN: A smart contract vulnerability detection method based on Dual Attention Graph Neural Network},
  author  = {Zhen, Zixian and Zhao, Xiangfu and Zhang, Jinkai and Wang, Yichen and Chen, Haiyue},
  journal = {Computer Networks},
  volume  = {242},
  pages   = {110238--110248},
  year    = {2024}
}

@inproceedings{10.1145/3543507.3583367,
  author    = {Qian, Peng and Liu, Zhenguang and Yin, Yifang and He, Qinming},
  title     = {Cross-Modality Mutual Learning for Enhancing Smart Contract Vulnerability Detection on Bytecode},
  year      = {2023},
  booktitle = {Proceedings of the ACM Web Conference (WWW)},
  pages     = {2220–2229},
  numpages  = {10}
}

@article{ma2025uncovering,
  author  = {Ma, Zuchao and Jiang, Muhui and Luo, Xiapu and Wang, Haoyu and Zhou, Yajin},
  journal = {IEEE Transactions on Dependable and Secure Computing},
  title   = {Uncovering NFT Domain-Specific Defects on Smart Contract Bytecode},
  year    = {2025},
  volume  = {22},
  number  = {5},
  pages   = {4877-4895}
}

@inproceedings{wang2023smart,
  author    = {Wang, Yichuan and Zhao, Jingjing and Zhang, Yaling and Hei, Xinhong and Zhu, Lei},
  title     = {Smart contract symbol execution vulnerability detection method based on {CFG} path pruning},
  booktitle = {Proceedings of the 5th {ACM} International Symposium on Blockchain and Secure Critical Infrastructure},
  year      = {2023},
  pages     = {132--139}
}

@article{10700860,
  author  = {Wang, Yichen and Zhao, Xiangfu and He, Long and Zhen, Zixian and Chen, Haiyue},
  journal = {IEEE Transactions on Network Science and Engineering},
  title   = {ContractGNN: Ethereum Smart Contract Vulnerability Detection Based on Vulnerability Sub-Graphs and Graph Neural Networks},
  year    = {2024},
  volume  = {11},
  number  = {6},
  pages   = {6382-6395}
}

@article{10.1145/3699597,
  author  = {Xiang, Jianhang and Gao, Zhipeng and Bao, Lingfeng and Hu, Xing and Chen, Jiayuan and Xia, Xin},
  title   = {Automating Comment Generation for Smart Contract from Bytecode},
  year    = {2025},
  volume  = {34},
  number  = {3},
  journal = {ACM Transactions on Software Engineering and Methodology},
  pages   = {1-31}
}

@article{10486822,
  author  = {Zheng, Zibin and Su, Jianzhong and Chen, Jiachi and Lo, David and Zhong, Zhijie and Ye, Mingxi},
  journal = {IEEE Transactions on Software Engineering},
  title   = {DAppSCAN: Building Large-Scale Datasets for Smart Contract Weaknesses in DApp Projects},
  year    = {2024},
  volume  = {50},
  number  = {6},
  pages   = {1360-1373}
}

@inproceedings{10646885,
  author    = {Wang, Sally Junsong and Pei, Kexin and Yang, Junfeng},
  booktitle = {Proceedings of the 2024 IEEE Symposium on Security and Privacy (SP)},
  title     = {SmartInv: Multimodal Learning for Smart Contract Invariant Inference},
  year      = {2024},
  volume    = {},
  number    = {},
  pages     = {2217-2235}
}

@article{LI2026122645,
  title   = {An attack detection mechanism in smart contracts based on deep learning and feature fusion},
  journal = {Information Sciences},
  volume  = {722},
  pages   = {122645--122664},
  year    = {2025},
  author  = {Peiqiang Li and Guojun Wang and Wanyi Gu and Xubin Li and Xiangyong Liu and Yuheng Zhang}
}


\end{document}